\documentclass[11pt,conference]{IEEEtran}
\usepackage{cite}
\usepackage{amsmath,amssymb,amsfonts}
\usepackage{algorithmic}
\usepackage{graphicx}
\usepackage{textcomp}
\usepackage{xcolor}
\usepackage[hyphens]{url}
\usepackage{fancyhdr}
\usepackage{hyperref}
\usepackage[utf8]{inputenc}

\usepackage{xspace}

\title{
DR-CGRA: Supporting Loop-Carried Dependencies in CGRAs Without Spilling Intermediate Values}

\author{Elad Hadar\\
    Technion - Israel Institute of Technology\\
    \texttt{ehadar@technion.ac.il}
    \and 
    Yoav Etsion\\
    Technion - Israel Institute of Technology\\
    \texttt{yetsion@technion.ac.il}
    }

\begin{document}
\maketitle

\begin{abstract}

Coarse-grain reconfigurable architectures (CGRAs) are gaining traction thanks to their performance and power efficiency. Utilizing CGRAs to accelerate the execution of tight loops holds great potential for achieving significant overall performance gains, as a substantial portion of program execution time is dedicated to tight loops. But loop parallelization using CGRAs is challenging because of loop-carried data dependencies. Traditionally, loop-carried dependencies are handled by spilling dependent values out of the reconfigurable array to a memory medium and then feeding them back to the grid. Spilling the values and feeding them back into the grid imposes additional latencies and logic that impede performance and limit parallelism. 

In this paper, we present the Dependency Resolved CGRA (DR-CGRA) architecture that is designed to accelerate the execution of tight loops. DR-CGRA, which is based on a massively-multithreaded CGRA, runs each iteration as a separate CGRA thread and maps loop-carried data dependencies to inter-thread communication inside the grid. This design ensures the passage of data-dependent values across loop iterations without spilling them out of the grid.

The proposed DR-CGRA architecture was evaluated on various SPEC CPU 2017 benchmarks. The results demonstrated significant performance improvements, with an average speedup ranging from 2.1 to 4.5 and an overall average of 3.1 when compared to state-of-the-art CGRA architecture.

\end{abstract}

\section{Introduction}

Coarse-grain reconfigurable architectures (CGRAs) have gained significant popularity across various fields, including high-performance computing (HPC), machine learning, and biomedicine \cite{kasgen2018coarse,kulkarni2016fully,chang2022reinforcement,mercado2023coarse,patel2011syscore,kim2014ulp}. The appeal of CGRAs lies in their ability to adapt the architecture to specific applications and their power efficiency in terms of performance per watt \cite{prabhakar2017plasticine}.

The rising popularity of CGRAs is further fueled by a slowdown in the performance advancements of general-purpose architectures in recent years \cite{theis2017end, chien2013moore,palem2012end,hameed2010understanding}. Consequently, reconfigurable architectures have been increasingly integrated into heterogeneous systems. These architectures leverage their diversity to achieve higher performance per watt across a broader range of applications compared to conventional general-purpose processors \cite{putnam2015reconfigurable}.

Sequential programs spend a considerable portion of their run time executing tight loops with loop-carried data dependencies, and accelerating these loops becomes a critical factor in enhancing overall performance \cite{wilhelm2008worst,cotterell2002synthesis,lee1999low,lee1999instruction,davidson1995aggressive,villarreal2001study}. The challenge in accelerating the execution of loops with loop-carried dependencies in CGRAs arises from the need to spill data-dependent values out of the array and then feed them back in \cite{govindaraju2011dynamically,govindaraju2012dyser,srinath2014architectural,gupta2011bundled}. Consequently, the execution of the next iteration is stalled until all its input operands become ready. The presence of this type of dependency in all loop iterations ultimately results in sequential execution, which leads to reduced performance \cite{vachharajani2007speculative}. We require an efficient mechanism to facilitate feedback of loop-dependent values between functional units, ensuring the data remains within the CGRA.

In this paper, we present the Dependency Resolved CGRA (DR-CGRA) architecture that is designed to accelerate the execution of tight loops with loop-carried data dependencies. DR-CGRA is based on the hybrid data-flow/von Neumann vector graph instruction word (VGIW) \cite{voitsechov2015control,voitsechov2014single} architecture. By extending the execution model, and the underlying hardware, the new architecture enables the passage of loop-carried dependent values across threads through the CGRA fabric. 

DR-CGRA leverages the multithreaded CGRA (MT-CGRA) core of the VGIW. The VGIW architecture introduced the multithreaded coarse-grained reconfigurable architecture array (MT-CGRA) core. The MT-CGRA core concurrently executes multiple instances of a control and data-flow graph, eliminating global pipeline overheads by directly communicating intermediate values across functional units. It also eliminates the need for a large register file. Furthermore, the underlying data-flow model enables the core to extract more instruction-level parallelism by allowing different types of functional units to execute in parallel.

DR-CGRA enables the efficient execution of tight loops with loop-carried data dependencies by treating each iteration as a separate thread. It maps backward edges, which represent loop-carried data dependencies, to inter-thread communication in the MT-CGRA core without spilling the communicated values out of the grid \cite{voitsechov2018inter}. The architecture manipulates the thread ID of the data going on the back edge, ensuring seamless data transfer. As a result, the DR-CGRA effectively eliminates the additional latencies and logic associated with traditional methods of handling loop-carried dependencies, optimizing performance.

In this paper, we make the following contributions:
\begin{itemize}
  \item We present the DR-CGRA execution model to run tight loops in a CGRA.
  \item We present the design of the DR-CGRA architecture.
  \item We evaluate the DR-CGRA architecture in simulation and show it provides an average speedup of 3.1 and up to 4.5.
\end{itemize}

The remainder of this paper is organized as follows: Section 2 introduces the execution model, providing a comprehensive overview. Following that, Section 3 presents a detailed description of the architecture. In Section 4, we delve into the methodology employed for this study. The obtained evaluation results are discussed in Section 5. Section 6 explores related work in the field. Finally, our findings are concluded in Section 7.

\section{Execution Model and loop-carried Dependency Patterns}

The execution model of the DR-CGRA is designed to efficiently accelerate the execution of tight loops in highly repetitive serial code. It addresses the challenges posed by loop-carried data dependencies, which significantly hamper the performance of various computational workloads, particularly in parallel computing. These dependencies emerge when subsequent iterations of a loop rely on the output of previous iterations. Consequently, they introduce a serialization constraint, compelling the loop to be executed sequentially, thereby hindering opportunities for parallelization.

DR-CGRA utilizes a massively-multithreaded CGRA design, wherein each iteration of a loop is executed as a separate CGRA thread. The simultaneous threads are collectively referred to as a "thread group," analogous to the concept of "warp" in GPUs. It is a specialized extension of VGIW designed to overcome its limitations in handling loop-carried data dependencies. While the VGIW architecture is explicitly parallel, relying on multiple threads to operate on different data, it faces constraints when a single thread needs to run a tight, dependent loop. This is due to the need to spill dependent values out of the grid and then feed them back in, causing the execution of the next iteration to stall until all input operands are ready. As a result, when this type of dependency occurs in all loop iterations, it leads to the sequential execution of threads associated with each iteration, reducing thread-level parallelism and significantly impacting performance.

To efficiently accelerate the execution of tight loops, DR-CGRA allows the mapping of dependent values across thread iterations directly inside the reconfigurable grid. It employs backward edges that enable the transfer of results from a compute unit back as input of that unit. It modifies the thread ID of the data going on the back edge and changes it to that of the corresponding next thread iteration ID. The data with a modified thread ID is then written to the existing token buffer of that unit. 
Additionally, the architecture facilitates data transfer through the original computational path when initial values are loaded or when there is no data dependency in the execution process.

The architecture's execution model effectively eliminates the additional latencies and logic associated with traditional methods of handling loop-carried dependencies, which ultimately impede performance and limit parallelism. By facilitating efficient data transfer within the array, DR-CGRA optimizes the execution of tight loops and efficiently handles loop-carried dependencies without stalling threads, thereby contributing to significant overall performance gains.

The execution model of the DR-CGRA incorporates a compilation stage to identify and analyze data dependencies within the code. This stage plays a crucial role in understanding the loop-carried data dependencies present in the serial code. During that stage, the code is analyzed to identify loops and their respective loop-carried dependencies.
Additionally, a "diff" value is determined, indicating the distance between the producer and consumer of a loop-carried variable in terms of the number of iterations. 
The primary objective of the analysis is to identify loop structures and determine the data dependencies between different iterations of each loop. It focuses on understanding how data is passed between iterations and also identifies the specific operand in which the dependent variable is involved in the operations. By identifying and analyzing these data dependencies, the DR-CGRA can accurately map loop-carried dependencies onto the reconfigurable array. The identified data dependencies are used to guide the mapping of loop-carried data dependencies onto the reconfigurable array.

\subsection{Data-dependent loop patterns}

\begin{figure}
  \includegraphics[width=7cm,height=3.5cm]{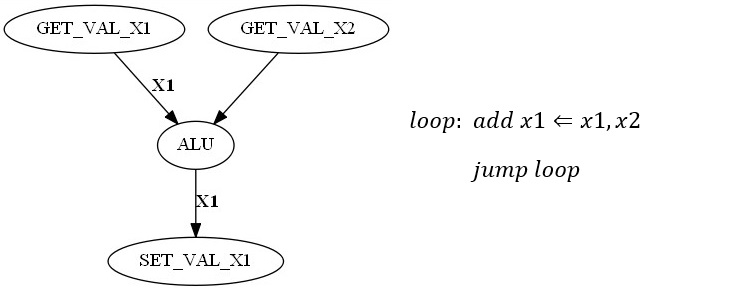}
  \caption{Simple, single-path dependency. Loop-carried data dependency where only units in the dependency path use that data-dependent variable. With no memory access\label{fig:scenario_1}}
\end{figure}

Various patterns of loop-carried data dependencies may arise, necessitating a thorough review when devising a data transfer solution. Analyzing and understanding these patterns are essential steps in developing efficient methods for handling loop-carried data dependencies in data-flow architecture designs. The possible patterns:

\textbf{Simple, single-path dependency (Figure~\ref{fig:scenario_1})}: Loop-carried data dependency when only the units within the dependent path are involved, and no memory access comes into play. This pattern scenario arises when the dependency solely affects the compute path, and other compute units in the grid do not consume the dependent variable either before or after its update. 
In this pattern, a compute operation updates the value of a certain variable. This operation is repeated since it is part of a loop. On every repeat of this operation, the updated result value must serve as the input of the same operation in the following iteration. 
Figure~\ref{fig:scenario_1} illustrates an example of this pattern, along with the corresponding pseudo-code.
For instance, in the provided pseudo code, the value of the input variable X1 is obtained from the previous iteration (Get\_val\_X1), and the updated output value of X1 is then set as an input for the subsequent iteration (Set\_Val\_X1).

\begin{figure}
  \includegraphics[width=7cm,height=4cm]{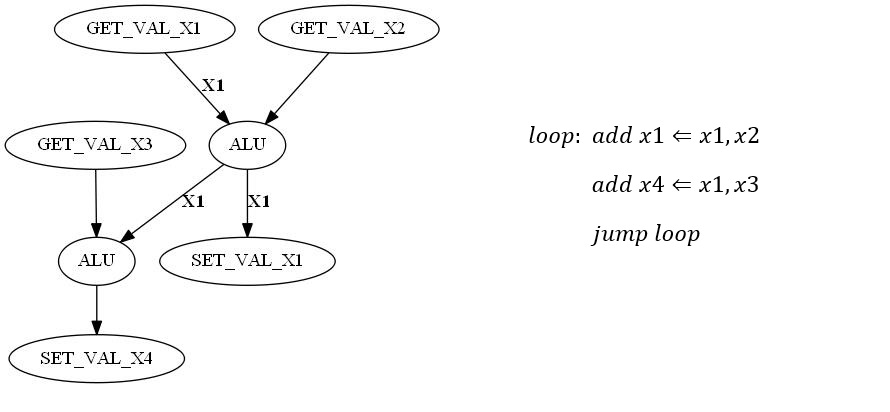}
  \caption{Dependency with diverging paths. Loop-carried data dependency where other units in the grid consume the data-dependent variable after the update of the data-dependent variable\label{fig:scenario_2}}
\end{figure}

\begin{figure}
  \includegraphics[width=7cm,height=4cm]{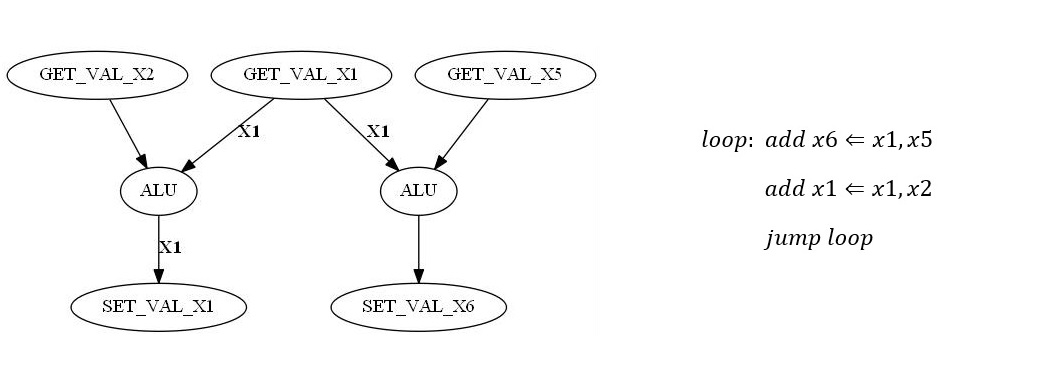}
  \caption{Dependency with diverging paths. Loop-carried data dependency where other units in the grid consume the data-dependent variable prior to the update of the data-dependent variable \label{fig:scenario_3}}
\end{figure}

\textbf{Dependency with diverging paths (Figures~\ref{fig:scenario_2} and~\ref{fig:scenario_3})}: Loop-carried data dependency that involves additional units within the loop iterations beyond those present in the dependent path and not involving memory access. In this pattern, these additional units only consume the produced dependent data without modifying it; otherwise, they would be considered part of the dependent path. Two scenarios arise in this pattern:
a) The other units in the grid consume the dependent variable after it has been updated. In this scenario, when resolving the data dependency issue, a straightforward connection between the units suffices, similar to regular cases.
b) The other units in the grid consume the dependent variable before it is updated. However, in this scenario, when resolving the data dependency issue locally within the compute operation over the dependent variable, it becomes crucial to ensure that the other units in the grid that consume the data before the update of that variable receive the correct value in the next iteration. Therefore, enabling updates of values at the end of the compute route becomes essential. This enables the next thread to reload the correct value every cycle during the next iterations, maintaining an uninterrupted flow of data. The reloading process introduces a constant delay, which depends on the distance between the update of that data-dependent value and the length of the data-flow compute path.

Figure~\ref{fig:scenario_2} demonstrates an example of this pattern with scenario A, where the data is consumed by other units in the grid after the update of the data-dependent variable. On the other hand, Figure~\ref{fig:scenario_3} depicts this pattern with scenario B, wherein the data is consumed by other units in the grid prior to the update of the data-dependent variable.

\begin{figure}
  \includegraphics[width=7cm,height=4cm]{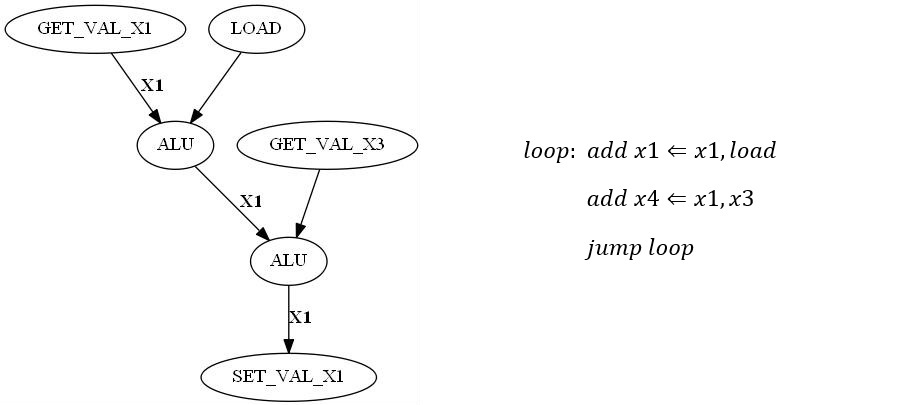}
  \caption{Dependency with background memory access. Loop-carried data dependency with memory access not on the dependent value\label{fig:scenario_4}}
\end{figure}

\textbf{Dependency with background memory access (Figure~\ref{fig:scenario_4})}: Loop-carried data dependency that involves memory access (not on the dependent value) is characterized by the presence of memory load or store operations within the dependent path or other paths in a loop. This introduces additional latency due to memory accesses, although multiple accesses can happen simultaneously, mitigating some of the delays. 
Figure~\ref{fig:scenario_4} demonstrates an example of this pattern, where the value of the input variable X1 is obtained from the previous iteration and added to a loaded variable from memory. 

\begin{figure}
  \includegraphics[width=7cm,height=6cm]{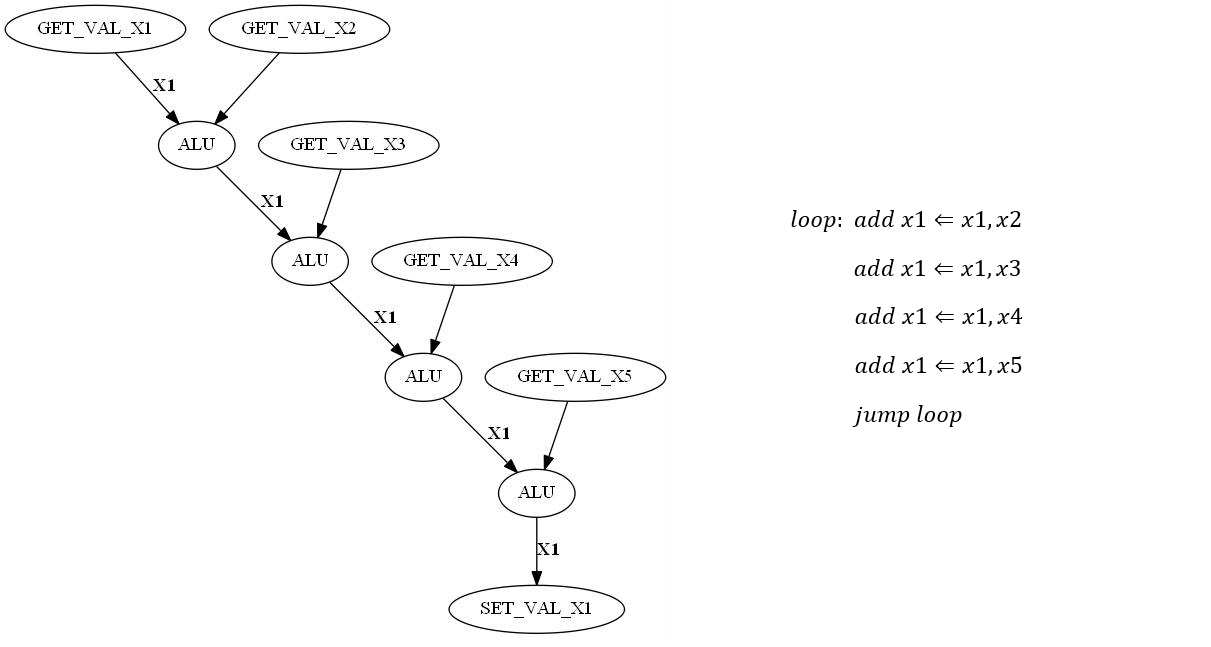}
  \caption{Consecutive Dependency. Loop-carried data dependency with consecutive loop-carried data dependency between units. The variable X1 is updated several times in each loop iteration \label{fig:scenario_5}}
\end{figure}

\textbf{Consecutive Dependency (Figure~\ref{fig:scenario_5})}: Consecutive loop-carried data dependency on the same variable occurs when the dependent path consists of several consecutive updates of the same variable, similar to the first pattern. Figure~\ref{fig:scenario_5} illustrates an example of this dependency pattern, where the data is consumed consecutively between the compute units.

The loop patterns illustrate different behaviors of data dependencies. In the following section, we will introduce the proposed DR-CGRA architecture, which supports the execution model and offers a solution for efficiently transferring data for various data dependency patterns.

\section{Architecture}

This section presents the DR-CGRA architecture, which extends the VGIW architecture to enable inter-thread transfer of loop-carried data dependencies.

\begin{figure}
  \includegraphics[width=7cm,height=7cm]{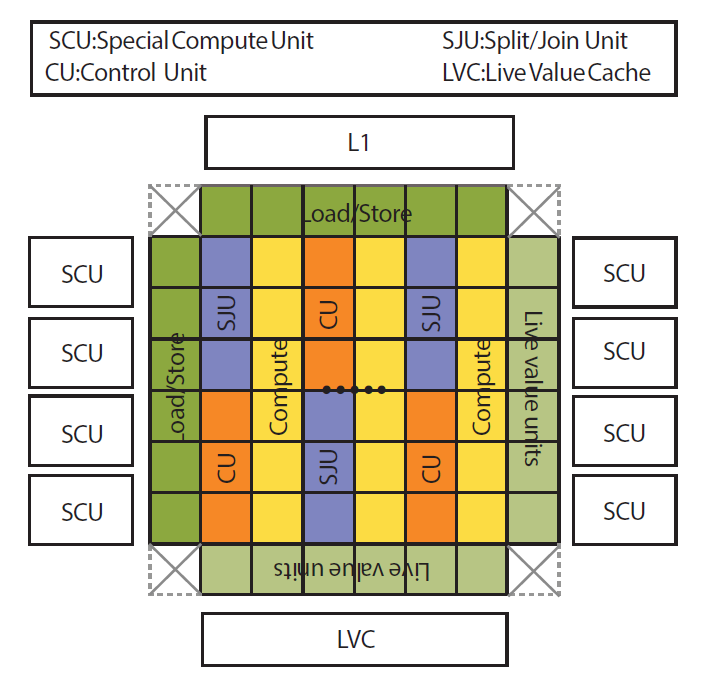}
  \caption{The VGIW MT-CGRA core \label{fig:arch_1}}
\end{figure}

The structure of the VGIW architecture is depicted in Figure~\ref{fig:arch_1}. The VGIW core comprises a grid of functional units interconnected by a statically routed network on chip (NoC). The core configuration, including the assignment of instructions to specific functional units and the routing scheme within the NoC, is determined during the compilation process and programmed into the reconfigurable array. During the execution phase, tokens are passed between the functional units in accordance with the predefined mapping established by the NoC. 
The grid itself is composed of heterogeneous functional units, each dedicated to specific types of operations. Memory operations are assigned to the load/store units; Computational operations find their place in the floating-point units and ALUs (compute units); Control operations are handled by the control units (CU), while split/join operations are responsible for preserving the original intra-thread memory order are carried out by the Split/Join units (SJU). 

\begin{figure*}
  \includegraphics[width=\textwidth,height=7cm]{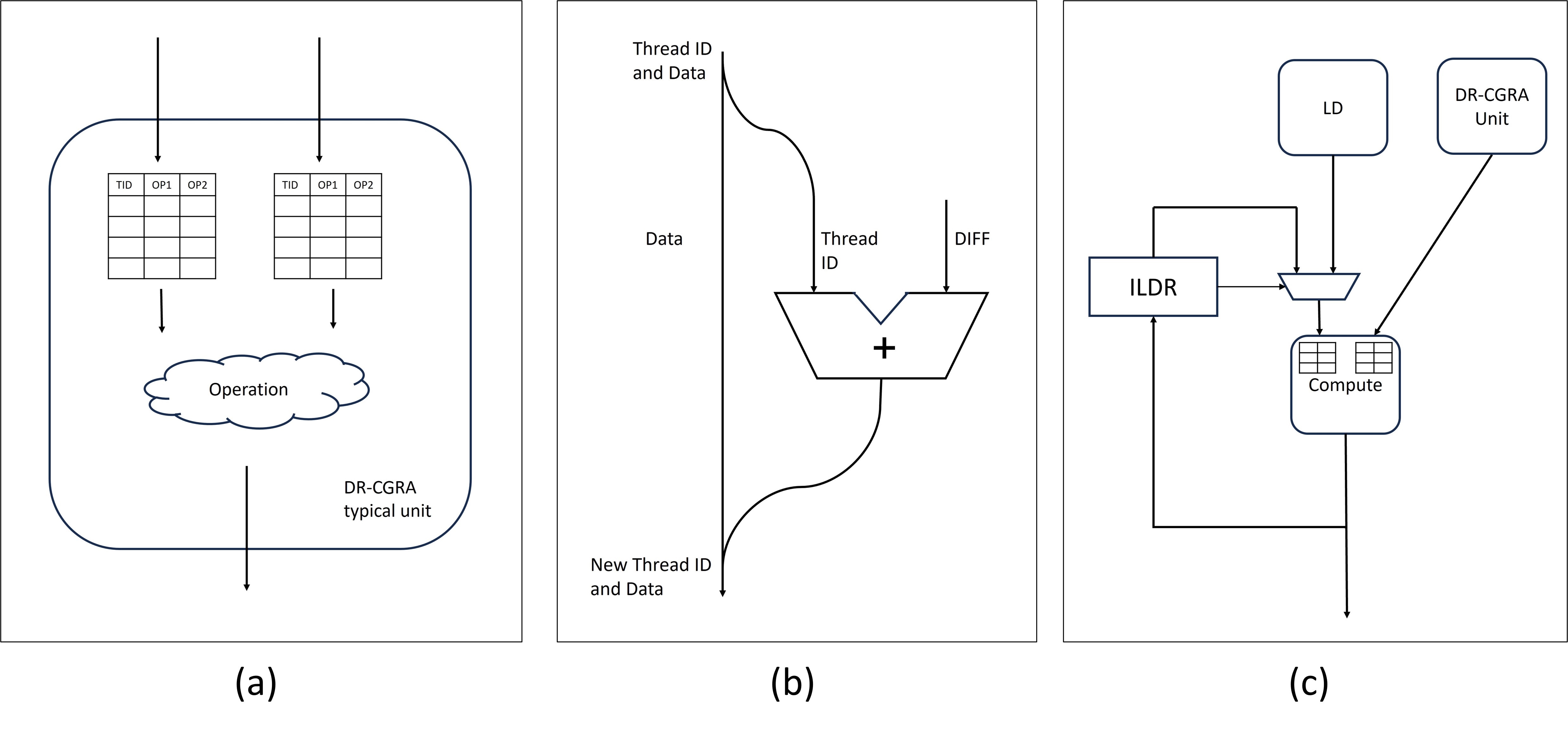}
  \caption{(a) shows a typical unit of the DR-CGRA, comprising a token buffer at each input. These token buffers serve to store the associated thread ID and operands until a matching thread ID of the two inputs is encountered. (b) illustrates the inner architecture of the ILDR, responsible for manipulating the thread ID value. (c) displays the integration of the ILDR and selector with each DR-CGRA compute unit. \label{fig:arch_5}}
\end{figure*}

During the execution of parallel tasks on the VGIW core, many distinct flows representing different threads coexist simultaneously within the grid. To enable effective information transfer, tagged tokens are composed of data and an associated Thread ID acting as a tag. The Thread ID serves as a crucial identifier, allowing the nodes in the grid to discern which operands belong to specific threads, ensuring accurate data handling.

The DR-CGRA introduces an addition to the VGIW architecture called the Inter-loop Dependency Resolution Unit (ILDR). Unlike existing units that solely manipulate the token data, the ILDR is designed to modify the tags associated with the tokens. This capability is crucial for loop-carried data transfer, as it allows for the necessary modifications to the tags of existing tokens, enabling efficient data transfer between loops. By introducing the ILDR, the architecture gains enhanced support for efficient loop-carried data transfer within the grid. This unit is added to the CGRA and attached to the compute units in the grid.

All of the DR-CGRA units incorporate tagged-token matching logic, which plays a vital role in facilitating thread interleaving through dynamic data-flow. As soon as all the required operands for a particular Thread ID become available, they are transferred to the respective unit's logic. Once the unit's operation is complete, the resulting output is transmitted back through the unit's switch, reintegrating it into the grid for further processing and coordination. Figure~\ref{fig:arch_5}a shows the common structure of the DR-CGRA units, featuring a token buffer at each input of the unit.

\subsection{Inter-loop Dependency Resolution Unit (ILDR)}

During the execution of loop iterations, the ILDR plays a critical role in facilitating the transfer of data from a compute unit that calculates dependent data. It ensures that the output of the compute unit in the current thread's execution path is fed back to the dependent operand input of the same unit in the next thread iteration's execution path.
Each compute unit within the reconfigurable grid is equipped with an ILDR along with a selector. The ILDR is responsible for transferring the output result from the compute unit in the current thread's execution path to the input of the same compute unit in the subsequent thread's execution path while updating the thread ID. The selector, in turn, determines whether to choose the original input loaded to the token buffer or the transferred data-dependent variable from the output of the compute unit.

The ILDR plays a crucial role in facilitating data transfer across thread iterations by updating the associated thread ID to reflect the next thread. This is accomplished by incorporating an adder that adds the thread ID with the "diff" value. By enabling this data transfer, the ILDR effectively reduces the latency between the output result and the input value to a single cycle. This eliminates the need for the consumer thread to wait for the producer thread to finish the entire execution path and avoids spilling the dependent value and unnecessary stalls.

Figure~\ref{fig:arch_5}c illustrates the feedback mechanism of the data-dependent variable across threads using the ILDR and selector, demonstrating the flow of data within the architecture.

The inner arch of the ILDR is depicted in Figure~\ref{fig:arch_5}b. The output result from the compute unit, as shown in Figure~\ref{fig:arch_5}c, continues uninterrupted to the input selector of the same compute unit. Simultaneously, the thread ID associated with the result data is redirected to the ILDR. Within the ILDR, the thread ID undergoes processing through an adder and is added to the "diff" value. The updated thread ID is then fed to the input selector of the compute unit. As a result, the compute unit efficiently writes the result into its token buffer, ensuring the data is readily available for the next thread's execution with the matching thread ID.

The DR-CGRA architecture inherently supports the transfer of a single data-dependent value across a compute unit. While extending support to two data-dependent input operands would require additional hardware resources, scenarios involving more than one dependent input variable are uncommon. Thus, the minimal impact on average performance gain does not justify incorporating extra hardware in such cases.

However, solely relying on the ILDR for feeding loop-carried dependent variables to subsequent units within the grid is insufficient. In certain cases, specific dependency paths within the grid rely on the variable's value earlier in the data-flow compute route. That is, the next iteration requires the input of a value calculated down the compute route (as described in section II - Loop patterns). Therefore, updating the value of the dependent variable at the end of the compute route becomes crucial to ensure proper propagation through the architecture. This approach guarantees that the data will be reloaded in the next thread, allowing it to reach the nodes requiring it as input.

\section{Experimental Methodology}

\subsection{Tested Benchmarks}

We evaluate the DR-CGRA architecture with benchmarks from the SPEC CPU 2017 benchmark suite \cite{SPEC17}. This benchmark suite offers a diverse range of compute-intensive workloads derived from actual user applications, enabling a comparative analysis of performance. Table I provides an overview of the specific benchmarks employed in the evaluation of the DR-CGRA architecture.

\begin{table*}[]
\centering
\resizebox{\textwidth}{!}{%
\begin{tabular}{lllll}
Benchmark & Name & Description &  &  \\\hline
500.perlbench\_r & Perl interpreter & cut-down version of Perl v5.22.1 &  &  \\
502.gcc\_r & GNU C compiler & based on GCC Version 4.5.0. It generates code   for an IA32 processor &  &  \\
505.mcf\_r & Route planning & a program used for single-depot vehicle   scheduling in public mass transportation &  &  \\
525.x264\_r & Video compression & application for encoding video streams into   the H.264/MPEG-4 AVC format &  &  \\
531.deepsjeng\_r & \begin{tabular}[c]{@{}l@{}}Artificial Intelligence: \\ alpha-beta tree   search (Chess)\end{tabular} & based on Deep Sjeng WC2008, the 2008 World   Computer Speed-Chess Champion &  &  \\
541.leela\_r & \begin{tabular}[c]{@{}l@{}}Artificial Intelligence: \\ Monte Carlo tree   search (Go)\end{tabular} & \begin{tabular}[c]{@{}l@{}}Go playing engine featuring Monte Carlo based   position estimation, selective \\ tree search based on Upper Confidence Bounds,   and move valuation based on Elo ratings.\end{tabular} &  &  \\
508.namd\_r & Molecular dynamics & \begin{tabular}[c]{@{}l@{}}derived from the data layout and inner loop   of NAMD, a parallel program for the \\ simulation of large biomolecular systems.\end{tabular} &  &  \\
511.povray\_r & Ray tracing & Computer Visualization / Raytracing &  &  \\
519.lbm\_r & Fluid dynamics & \begin{tabular}[c]{@{}l@{}}This program implements the so-called   "Lattice Boltzmann Method" (LBM) to \\ simulate incompressible fluids   in 3D\end{tabular} &  &  \\
521.wrf\_r & Weather forecasting & based on the Weather Research and Forecasting   Model (WRF) &  &  \\
538.imagick\_r & Image manipulation & ImageMagick is a software suite to create,   edit, compose, or convert bitmap images &  & 
\end{tabular}%
}
\caption{\label: Benchmarks}
\end{table*}

\subsection{Run time flow analysis}

In order to reduce simulation time, we wanted to focus on the loops that consumed the largest fraction of execution run time. To this end, we analyzed the program control flow and conducted a comprehensive run-time flow analysis on the SPEC 2017 benchmarks listed in Table I. This analysis involved the development of an Intel pin-tool program, which serves as a dynamic binary instrumentation tool \cite{Intel_pin,luk2005pin}. The pin-tool program monitored the execution of each benchmark, documenting the instructions, basic blocks, and routines involved. All benchmarks were compiled to generate standalone executable files.
The Analysis program provides reports on all the connections between basic blocks within each routine in every benchmark. We designed the program to consider basic blocks as containing a single jump instruction, enabling the identification of complete loop routes across all routines.

Within each benchmark, the routines underwent a detailed analysis to identify loop sections, particularly focusing on the top routines responsible for over 1\% of the total run time. To facilitate this process, a depth-first search (DFS) based algorithm was developed, enabling the identification of all basic block routes within these routines and determining the prevalence of each route. The algorithm specifically targeted closed routes representing loops that could be efficiently mapped to the reconfigurable grid using a single grid mapping. Subsequently, these identified routines and their corresponding loop sections were mapped to the DR-CGRA and then evaluated.

\subsection{DR-CGRA Performance}
We examined the speedup using a two-step evaluation process. Firstly, we mapped the basic block loop routes of each common routine in a specific benchmark onto the DR-CGRA. Then, we compared the run time of that section under two modes: a) The VGIW version without support for data transfer of loop-carried data dependency across iterations, which necessitates stalling instructions that consume the data-dependent variable until it becomes available. b) The DR-CGRA version that facilitates data transfer of loop-carried data dependency across iterations and threads. Subsequently, we calculated the average speedup of the accelerated parts while considering the prevalence of each loop route at run time. This comprehensive evaluation approach enabled us to quantify the performance gains achieved through tight loop acceleration.

\section{Evaluation}

This section presents our analysis of the benchmark tested with respect to their tight loop characteristics. Additionally, we explore the evaluation of the DR-CGRA architecture.

\subsection{Run time Evaluation }

\begin{figure}
  \includegraphics[width=\columnwidth,height=5cm]{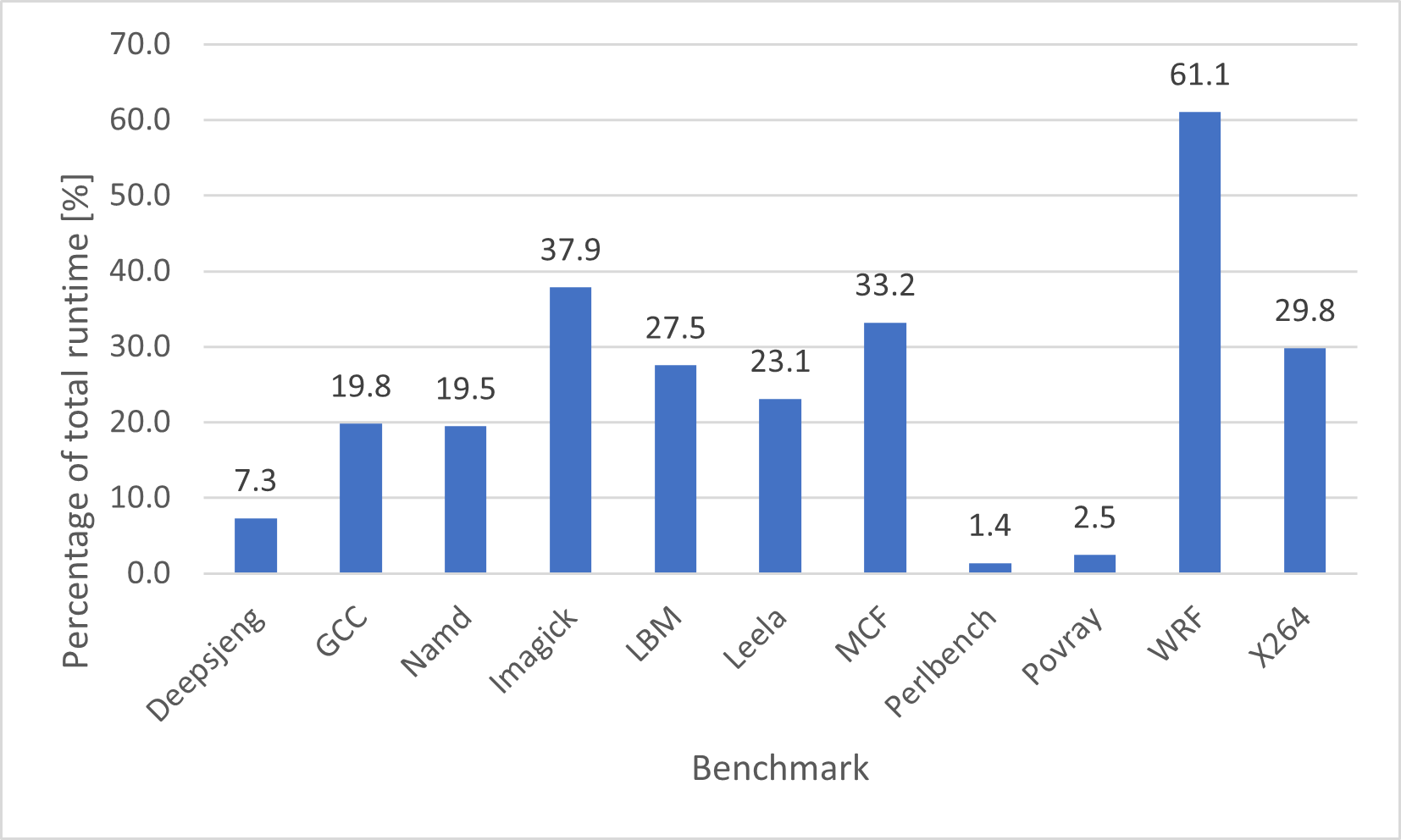}
  \caption{Percentage of total run time spent in tight loops for SPEC CPU 2017 benchmarks \label{fig:f_7}}
\end{figure}

Figure~\ref{fig:f_7} illustrates the average duration of benchmarks from the SPEC CPU 2017 suite spent within tight loops suitable for acceleration in the DR-CGRA architecture. The average run time was computed by considering all the instruction and basic block routes that form loops within the routines of each benchmark. Specifically, we focused on routines with a run time exceeding 1\% of the total benchmark run time. The analysis reveals that the time spent within tight loops across the benchmarks varies significantly, ranging from 1.4\% to 61.1\% with an average of 23.9 of the overall benchmark run time. These findings highlight the importance of optimizing loop execution for enhancing performance in these benchmarks.

\begin{figure*}
  \includegraphics[width=\textwidth,height=6cm]{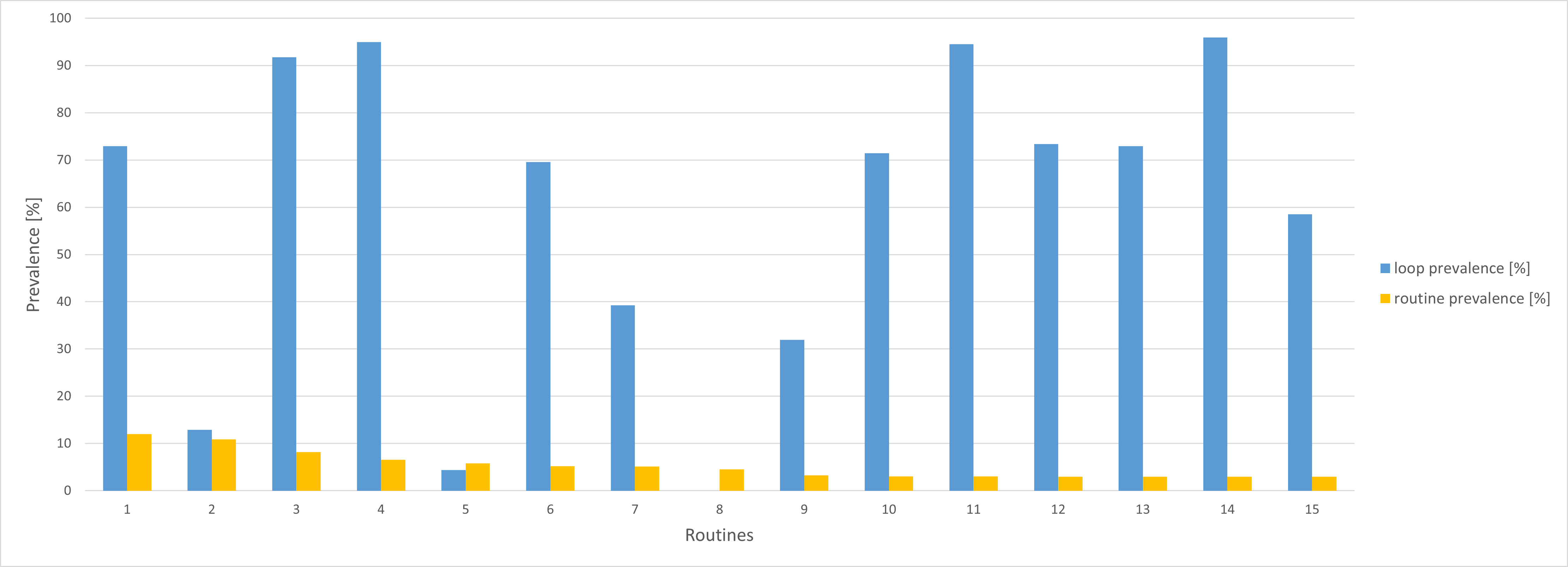}
  \caption{Routines prevalence and their respective percentage of time spent in loops \label{fig:f_8}}
\end{figure*}

The average time spent within loops is derived directly from the individual time each routine of the benchmark spends executing loops. It is important to note that the extent to which each routine is engaged in loops can vary significantly. The prevalence of a routine, relative to all other routines within the benchmark, often plays a crucial role in determining the average time spent in loops for the benchmark as a whole. Understanding these characteristics of the benchmarks aids in identifying the specific sections that can benefit from acceleration using the DR-CGRA architecture. Figure~\ref{fig:f_8} provides insights into the prevalence of routines in the WRF benchmark, as well as the prevalence of loops within each routine. Routine prevalence refers to the run time contribution of a routine to the tested routines' overall run time, while loop prevalence within a routine indicates the relative portion of time spent in loops within that routine. These metrics contribute to a deeper understanding of the benchmark's behavior and guide optimization efforts.

The analysis further shows that achieving acceleration for the majority of run time loop iterations in a specific application can be accomplished by focusing on accelerating only a small portion of the loops. This finding suggests that targeting specific critical tight loops can yield significant performance gains.

\begin{figure}
  \includegraphics[width=\columnwidth,height=5cm]{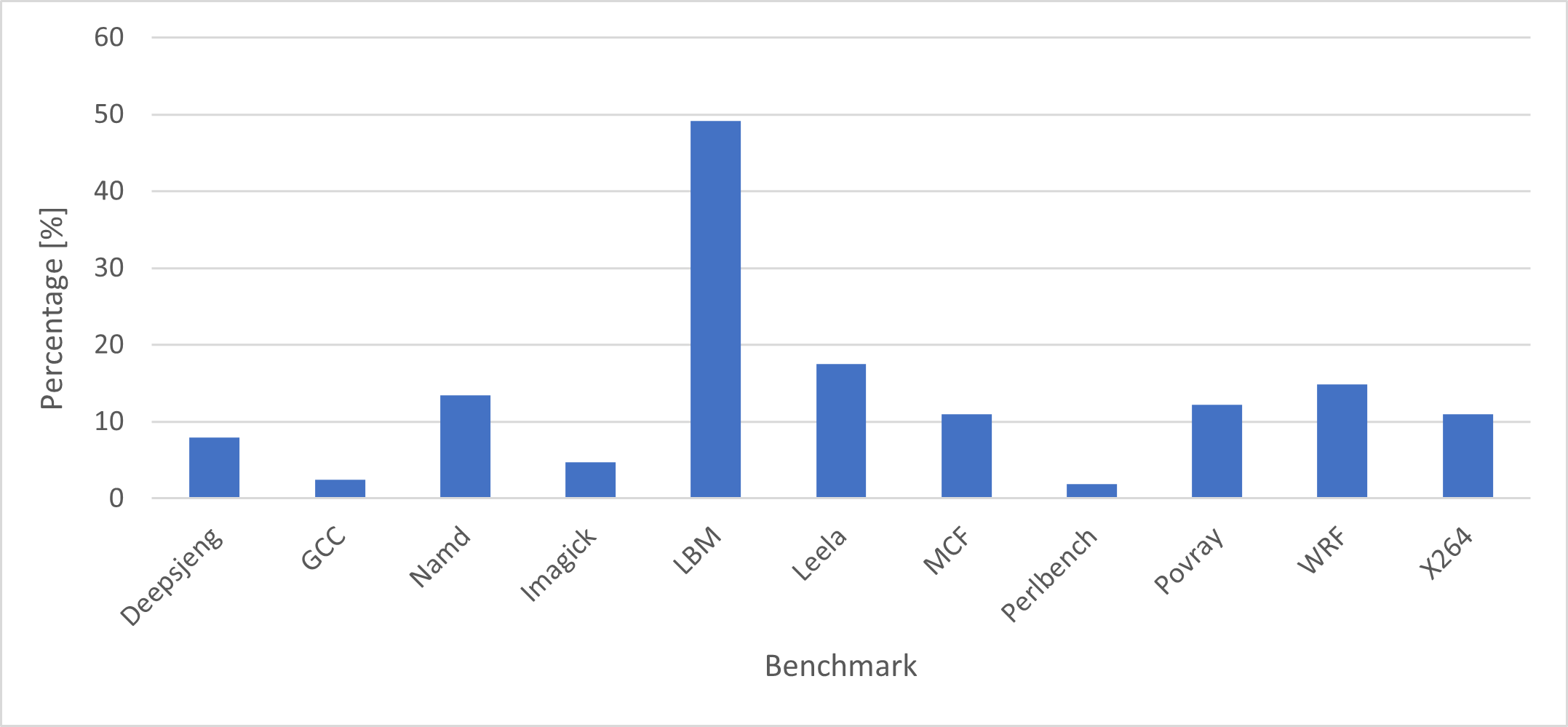}
  \caption{The portion of the most common loops that contribute to at least 90\% of the total run time of all iterations in a specific benchmark \label{fig:f_9}}
\end{figure}

In Figure~\ref{fig:f_9}, the analysis reveals the percentage of the most common loops that contribute to at least 90\% of the total run time iterations in a specific benchmark. Among the 11 benchmarks, it is evident that in 10 cases, less than 20\% of the loops account for 90\% of the entire loop iterations run time. On average, across all tested benchmarks, approximately 13.33\% of the loops are responsible for 90\% of the total loop iteration, and an average of 18.3\% of all loops accounting for 95\% of the entire loop iterations. These findings underscore the significance of focusing on optimizing the critical loops to achieve substantial performance improvements.

In addition to assessing loop prevalence, our run-time flow analysis examined the prevalence of different loop-carried data dependency patterns. Interestingly, we found that the pattern illustrating consecutive loop-carried data dependency over the same variable is rare (loop pattern detailed in Section II). Consequently, no dedicated architectural solution was employed to address this specific pattern.

We omit the Perlbench, Povray, and Deepsjeng benchmarks from further DR-CGRA performance evaluation. The analysis revealed a low potential for overall performance gain in these benchmarks, mainly because the prevalence of tight loop sections suitable for acceleration was relatively low across the entire benchmark. As a result, focusing on accelerating these benchmarks would not yield significant improvements in overall performance.

Once we analyzed the run time behavior of the benchmarks and explored the potential for loop acceleration, we proceeded to evaluate the performance of the accelerated sections using the DR-CGRA architecture.

\subsection{DR-CGRA Performance Evaluation}

\begin{figure*}
  \includegraphics[width=\textwidth,height=6cm]{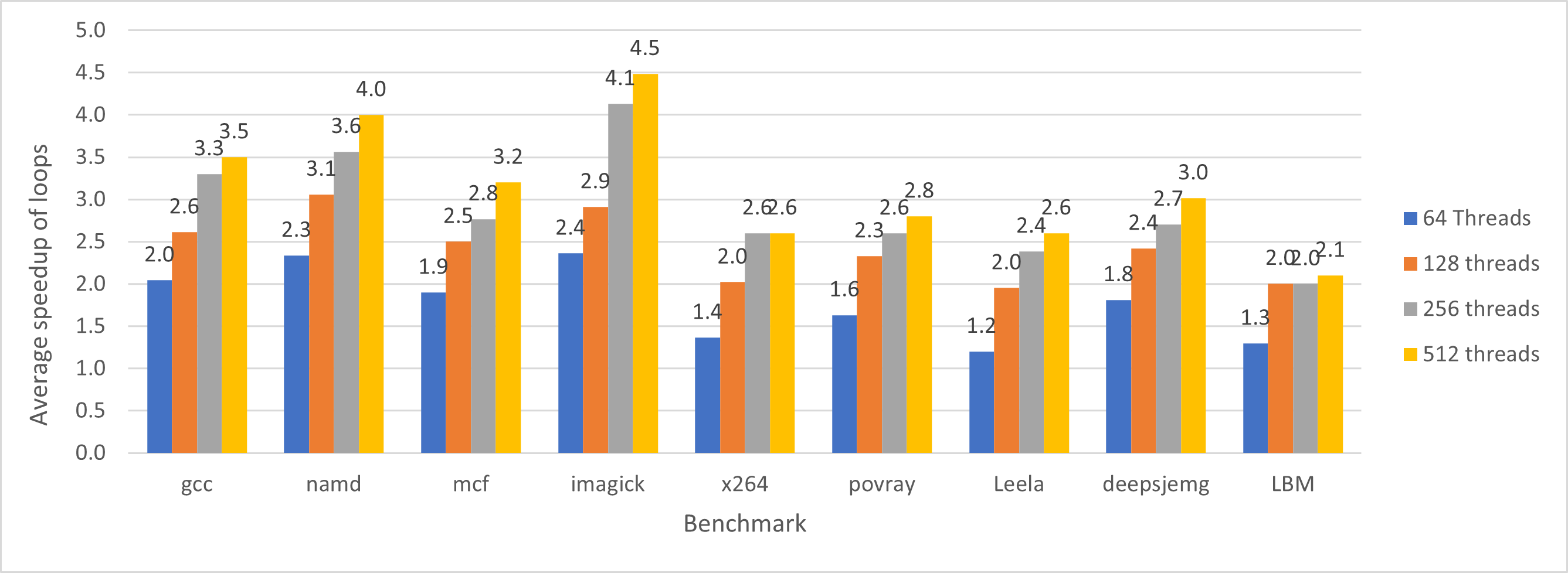}
  \caption{Average speedup achieved in each benchmark's accelerated loop sections when varying the number of threads \label{fig:f_16}}
\end{figure*}

As shown in Figure~\ref{fig:f_16}, the results of our evaluation demonstrate that the average speedup achieved for each benchmark's accelerated loop sections varies from 2.1 (for LBM) to 4.5 (for Imagick), and an overall
average of 3.1 when enabling an active thread group of 512 threads. This variation in speedup is primarily influenced by the instructions of the loop sections. The data-flow structure and data dependency characteristics are determined by the instructions within the loop route. Longer data-dependent paths within the loop lead to more stall cycles between iterations, making the speedup more significant when eliminating these stalls. In the case of the Imagick benchmark, the speedup was highest for its loop sections due to its most common loops in the two most common routines having long data-dependent paths. These two data-dependent routes account for over 60\% of all Imagick's loop iterations. On the other hand, in the LBM benchmark, most loops have very few instructions. Furthermore, most loops include memory access, which inhibits significant speedup. For most other benchmarks, the run time loop iterations are distributed between more loop sections, varying in the length of the data-dependent path and exhibiting different memory access patterns. These factors contribute to the observed range of speedup values across the benchmarks.

The research findings also reveal a significant influence of the thread group size on loop acceleration. As the number of threads within a thread group increases, the performance of tight loop execution also improves. The heightened thread-level parallelism, resulting from more active threads, greatly enhances overall system performance. This parallelism enables threads to execute simultaneously, effectively concealing computation and memory access latencies.

\begin{figure}
  \includegraphics[width=\columnwidth,height=4cm]{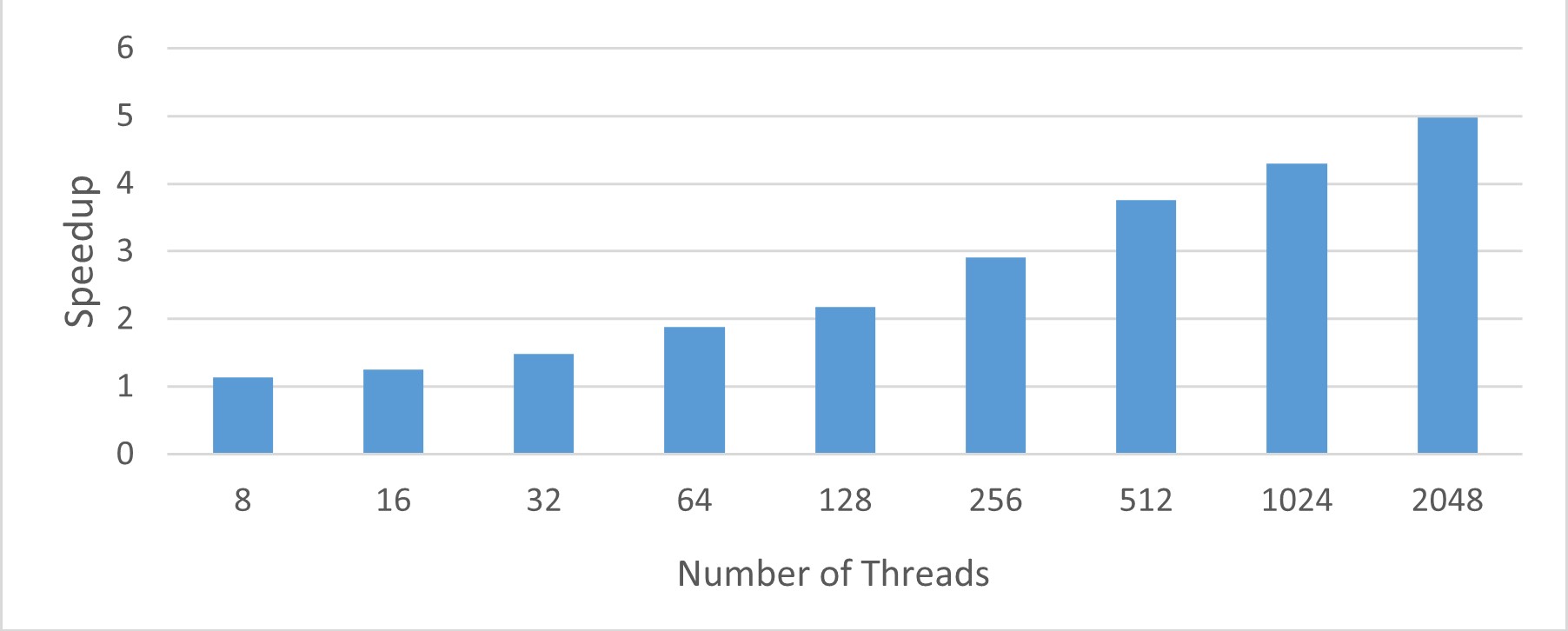}
  \caption{Speedup of a single accelerated loop section with memory access, obtained from the WRF benchmark, when varying the number of threads \label{fig:f_10}}
\end{figure}

\begin{figure}
  \includegraphics[width=\columnwidth,height=4cm]{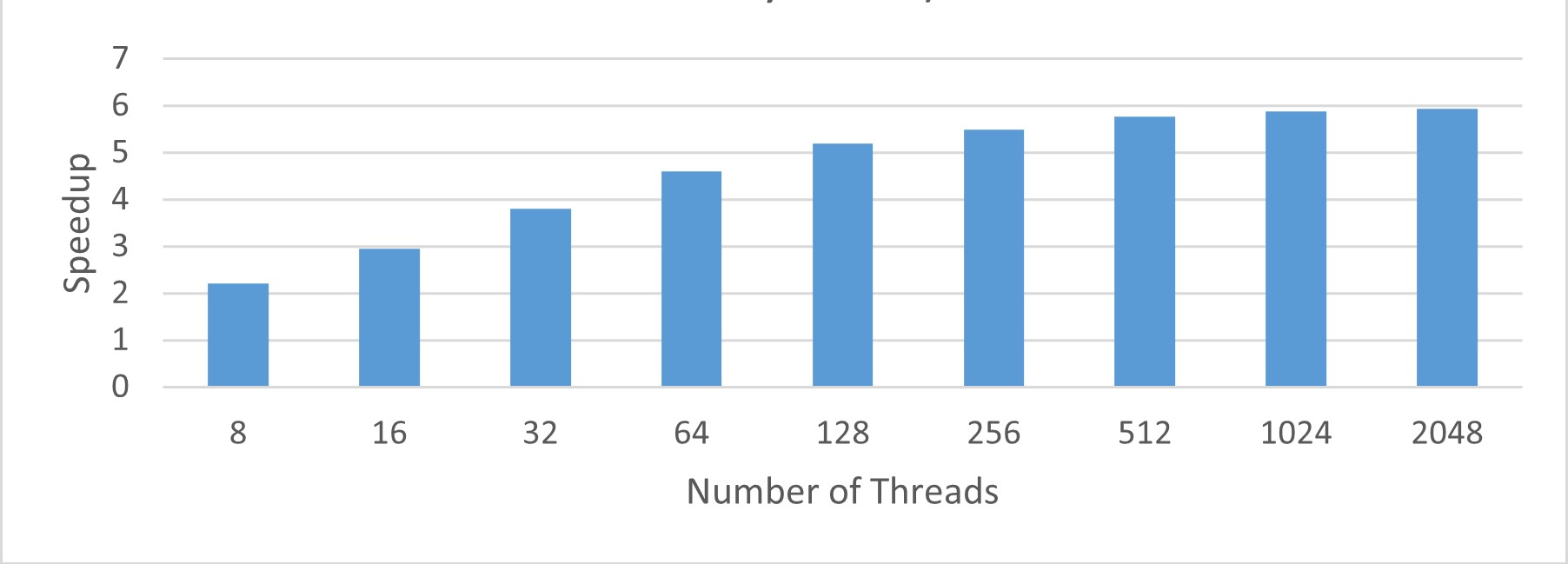}
  \caption{Speedup of a single accelerated loop section with no memory access, obtained from the WRF benchmark, when varying the number of threads \label{fig:f_11}}
\end{figure}

Figures~\ref{fig:f_10} and~\ref{fig:f_11} depict the performance speedup of a specific accelerated loop section from the WRF benchmark, observed with varying thread configurations. In Figure~\ref{fig:f_10}, the loop section's data-dependent path involves memory access, although not directly on the data-dependent variable. This means that the operation needs to wait for the data to arrive from memory, but it won't stall on every data input due to concurrent data fetching for subsequent threads (loop pattern detailed in Section II).
On the other hand, Figure~\ref{fig:f_11} illustrates the same loop section, with one notable change: the data-dependent path within the loop does not involve memory access. Comparing the two scenarios allows us to analyze the impact of memory access on the performance speedup.

The findings reveal that for both scenarios, with and without memory access within the loop, increasing the number of threads in the thread group leads to improved performance (as was also shown earlier). However, the performance increase is more significant for the case without memory access, particularly when using a smaller number of threads. For instance, the results demonstrate that the performance achieved with no memory access and 32 threads surpasses the performance obtained with memory access and 512 threads.

As sufficient threads are enabled, we observe that both cases experience an increase in speedup. However, it becomes evident that the rate of increase gradually slows down and approaches a certain point. Interestingly, the speedup values at that point are nearing a somewhat similar value for both cases.
This behavior is expected because a higher number of threads allows for better utilization of available parallelism, leading to improved performance. The reduction in latencies plays a crucial role in achieving this performance gain. These latencies include the fixed latency associated with the initial data insertion into critical path units mapped to the grid (often referred to as "filling the pipe") and the fixed latency of stalls inserted after the first thread in the data-dependent path until it reaches the compute unit and ILDR for the first time. These reductions significantly contribute to the observed performance improvement.
The advantage of the case without memory access is attributed to the additional latencies incurred in accessing memory for the case with memory access. As a result, the performance increase in the memory access scenario is less pronounced.
Moreover, the results demonstrate that when a sufficient number of threads are enabled, the speedup approaches six times the performance compared to scenarios without data transfer. This substantial speedup is achieved by eliminating stalls that occur between iterations due to the few-cycle-long data-dependent path in this specific loop. The path exists between the last unit outputting the dependent data and the first unit consuming it. By enabling data transfer, these stalls are eliminated, and all threads collaborate in covering the aforementioned latencies, thereby enhancing overall performance. 
These results shed light on how memory access influences the efficiency of the accelerated loop section in a multi-threaded execution environment.


In conclusion, our evaluation showcases the performance advantages of the DR-CGRA architecture compared to the VGIW MT-CGRA when executing loops with loop-carried data dependencies.

\section{Related Work}

Previous research has been driven by the potential performance benefits attainable through loop execution time acceleration. In pursuit of these objectives, various studies have been conducted, focusing on the following areas:

The Garp compiler and architecture \cite{hauser1997garp,callahan2000garp,callahan2000adapting} are specifically designed to facilitate the pipelined execution of loops on a co-processor. This system comprises a single-issue microprocessor equipped with a rapidly reconfigurable array acting as the co-processor for loop acceleration. The architecture employs a single thread that alternates between the microprocessor and the co-processor (array). During the activation of the co-processor, the microprocessor enters a sleep state, and upon loop completion, it resumes operation. 
To handle loop-carried data dependencies, Garp employs a mapping approach that aligns the loops to the array with their back edges. Resolving data dependencies is accomplished using a method akin to the static single assignment (SSA) form, incorporating immediate representation. Furthermore, the datapath of Garp incorporates Hold modules for each loop-carried variable input, which securely holds the dependent data.
While Garp shares some execution characteristics, including data dependency recognition, it does not support multiple threads, setting it apart from the present architecture.

The explicit loop specialization (XLOOPS) \cite{srinath2014architectural} is centered around encoding inter-iteration loop dependence patterns directly in the instruction set. It offers support for a wide range of inter-iteration data dependence patterns, necessitating changes to a general-purpose compiler that utilizes programmer annotations to generate XLOOPS binaries.
The core idea behind XLOOPS lies in expressing inter-iteration data dependence patterns through the use of specialized XLOOPS instructions. The XLOOPS microarchitecture enhances a general-purpose processor by incorporating a loop-pattern specialization unit. This unit consists of four decoupled lanes, each responsible for executing one iteration of a loop. The architecture efficiently resolves data dependencies by enabling data-dependent variables, identified through annotations, to be communicated across these four lanes. This facilitates the transmission of essential data to any other lane that requires it for the execution of its own iteration.
XLOOPS employs different approaches concerning data dependency recognition and data-dependent variable transfer. Within the architecture, the data-dependent variable is transferred by sharing the cross-iteration buffer among the lanes, which concurrently execute different iterations.

BERET \cite{gupta2011bundled} (Bundled Execution of REcurring Traces) is a configurable co-processor designed for handling recurring instruction sequences. The primary emphasis of BERET lies in optimizing short program regions that exhibit looping behavior. These regions' instruction sequences are disassembled into data-flow subgraphs, which are then efficiently mapped onto a set of subgraph execution blocks.
In its operation, BERET utilizes a single thread and incorporates an internal register file shared among the subgraphs.

DYSER (Dynamically specializing execution resources) \cite{govindaraju2011dynamically,govindaraju2012dyser} dynamically generates specialized data paths solely for frequently executed regions. This technique incorporates a heterogeneous array of functional units interconnected with simple switches, enabling efficient functionality and parallelism mechanisms.
When executing loop sections of the code, DYSER employs loop unrolling techniques to optimize performance. Moreover, to effectively handle loop-carried dependencies, the architecture facilitates loop parallelization by providing temporary storage for dependent variables.

Another proposal to accelerate loop run time using a specialized loop hardware accelerator is exemplified by the work of Fan et al. in the development of the loop accelerator (LA) \cite{fan2009bridging,fan2005cost}.
The LA presents a dedicated hardware implementation for scheduling loops. Each individual LA unit comprises a specialized datapath featuring tailored functional units, register files, an interconnect, and a simple controller.
In this architecture, each functional unit performs a specific set of functions customized for the particular loop it serves. To handle loop-carried data dependencies efficiently, the LA employs a strategy wherein each functional unit writes to a dedicated shift register file. The contents of these registers are then shifted downward to the next register in a pipeline fashion. Wires connecting the registers back to the functional unit inputs facilitate consistent data transfer from producers to consumers. Additionally, the LA allows direct point-to-point connectivity between functional units.

\section{Conclusions}
We introduced the Dependency Resolved CGRA (DR-CGRA) architecture, a novel extension of the VGIW that significantly enhances the performance of repetitive serial code, specifically tight loops. By addressing loop-carried data dependencies, we facilitated architectural changes within the reconfigurable grid, resulting in improved acceleration. The key innovation was the incorporation of the ILDR, which is responsible for efficiently transferring loop-carried dependencies between iterations. This architectural enhancement enabled the DR-CGRA to effectively handle loop-carried data dependencies. Furthermore, to support the loop analysis and optimization process, we developed a run-time analysis tool. This tool was designed to identify loops within the code, providing crucial insights that guided the architectural changes and optimizations performed by the DR-CGRA.
Through evaluation using the SPEC 2017 benchmark suite, we observed substantial performance gains in the accelerated loop sections. The average speedup ranged from 2.1 to 4.5, and an overall average of 3.1 when compared to state-of-the-art CGRA architecture, demonstrating the potential of the DR-CGRA architecture in optimizing loop execution.
Furthermore, our findings revealed that the majority of the tested benchmarks dedicate a substantial portion of their run time to executing tight loops.

\section*{Acknowledgements}
%We thank the anonymous referees for their valuable
%comments and suggestions. %This research was funded by the ... , Author 1, was supported by ...; 
E. Hadar was supported by the Hasso Plattner Institute (HPI).

\bibliographystyle{ieeetr}
\bibliography{refs}

\begin{thebibliography}{10}

\bibitem{kasgen2018coarse}
P.~S. K{\"a}sgen, M.~Weinhardt, and C.~Hochberger, ``A coarse-grained reconfigurable array for high-performance computing applications,'' in {\em 2018 Intl. Conf. on ReConFigurable Computing and FPGAs (ReConFig)}, IEEE, 2018.

\bibitem{kulkarni2016fully}
A.~Kulkarni, E.~Vasteenkiste, D.~Stroobandt, A.~Brokalakis, and A.~Nikitakis, ``A fully parameterized virtual coarse grained reconfigurable array for high performance computing applications,'' in {\em 2016 IEEE Intl. Parallel and Distributed Processing Symp. Workshops (IPDPSW)}, IEEE, 2016.

\bibitem{chang2022reinforcement}
A.~X.~M. Chang, P.~Khopkar, B.~Romanous, A.~Chaurasia, P.~Estep, S.~Windh, D.~Vanesko, S.~D.~B. Mohideen, and E.~Culurciello, ``Reinforcement learning approach for mapping applications to dataflow-based coarse-grained reconfigurable array,'' {\em arXiv preprint arXiv:2205.13675}, 2022.

\bibitem{mercado2023coarse}
K.~Mercado, S.~Bavikadi, and S.~M. PD, ``Coarse-grained high-speed reconfigurable array-based approximate accelerator for deep learning applications,'' in {\em 2023 57th Annual Conf. on Information Sciences and Systems (CISS)}, IEEE, 2023.

\bibitem{patel2011syscore}
K.~Patel, S.~McGettrick, and C.~J. Bleakley, ``Syscore: A coarse grained reconfigurable array architecture for low energy biosignal processing,'' in {\em 2011 IEEE 19th Annual Intl. Symp. on Field-Programmable Custom Computing Machines}, IEEE, 2011.

\bibitem{kim2014ulp}
C.~Kim, M.~Chung, Y.~Cho, M.~Konijnenburg, S.~Ryu, and J.~Kim, ``Ulp-srp: Ultra low-power samsung reconfigurable processor for biomedical applications,'' {\em ACM Transactions on Reconfigurable Technology and Systems (TRETS)}, vol.~7, no.~3, 2014.

\bibitem{prabhakar2017plasticine}
R.~Prabhakar, Y.~Zhang, D.~Koeplinger, M.~Feldman, T.~Zhao, S.~Hadjis, A.~Pedram, C.~Kozyrakis, and K.~Olukotun, ``Plasticine: A reconfigurable architecture for parallel paterns,'' {\em ACM SIGARCH Computer Architecture News}, vol.~45, no.~2, pp.~389--402, 2017.

\bibitem{theis2017end}
T.~N. Theis and H.-S.~P. Wong, ``The end of moore's law: A new beginning for information technology,'' {\em Computing in Science \& Engineering}, vol.~19, no.~2, 2017.

\bibitem{chien2013moore}
A.~A. Chien and V.~Karamcheti, ``Moore's law: The first ending and a new beginning,'' {\em Computer}, vol.~46, no.~12, 2013.

\bibitem{palem2012end}
K.~Palem and A.~Lingamneni, ``What to do about the end of moore's law, probably!,'' in {\em Proceedings of the 49th Annual Design Automation conf.}, 2012.

\bibitem{hameed2010understanding}
R.~Hameed, W.~Qadeer, M.~Wachs, O.~Azizi, A.~Solomatnikov, B.~C. Lee, S.~Richardson, C.~Kozyrakis, and M.~Horowitz, ``Understanding sources of inefficiency in general-purpose chips,'' in {\em Proceedings of the 37th annual Intl. Symp. on Computer architecture}, 2010.

\bibitem{putnam2015reconfigurable}
A.~Putnam, A.~M. Caulfield, E.~S. Chung, D.~Chiou, K.~Constantinides, J.~Demme, H.~Esmaeilzadeh, J.~Fowers, G.~P. Gopal, J.~Gray, {\em et~al.}, ``A reconfigurable fabric for accelerating large-scale datacenter services,'' {\em IEEE Micro}, vol.~35, no.~3, 2015.

\bibitem{wilhelm2008worst}
R.~Wilhelm, J.~Engblom, A.~Ermedahl, N.~Holsti, S.~Thesing, D.~Whalley, G.~Bernat, C.~Ferdinand, R.~Heckmann, T.~Mitra, {\em et~al.}, ``The worst-case execution-time problem—overview of methods and survey of tools,'' {\em ACM Transactions on Embedded Computing Systems (TECS)}, vol.~7, no.~3, 2008.

\bibitem{cotterell2002synthesis}
S.~Cotterell and F.~Vahid, ``Synthesis of customized loop caches for core-based embedded systems,'' in {\em Proceedings of the 2002 IEEE/ACM Intl. Conf. on Computer-aided design}, 2002.

\bibitem{lee1999low}
L.~H. Lee, B.~Moyer, J.~Arends, and A.~Arbor, ``Low-cost embedded program loop caching-revisited,'' {\em University of Michigan Technical Report CSE-TR-411-99}, 1999.

\bibitem{lee1999instruction}
L.~H. Lee, B.~Moyer, and J.~Arends, ``Instruction fetch energy reduction using loop caches for embedded applications with small tight loops,'' in {\em Proceedings of the 1999 Intl. Symp. on Low power electronics and design}, 1999.

\bibitem{davidson1995aggressive}
J.~W. Davidson and S.~Jinturkar, ``An aggressive approach to loop unrolling,'' tech. rep., Citeseer, 1995.

\bibitem{villarreal2001study}
J.~Villarreal, R.~Lysecky, S.~Cotterell, and F.~Vahid, ``A study on the loop behavior of embedded programs,'' {\em University of California, Riverside, Tech. Rep. UCR-CSE-01-03}, 2001.

\bibitem{govindaraju2011dynamically}
V.~Govindaraju, C.-H. Ho, and K.~Sankaralingam, ``Dynamically specialized datapaths for energy efficient computing,'' in {\em 2011 IEEE 17th Intl. Symp. on High Performance Computer Architecture}, IEEE, 2011.

\bibitem{govindaraju2012dyser}
V.~Govindaraju, C.-H. Ho, T.~Nowatzki, J.~Chhugani, N.~Satish, K.~Sankaralingam, and C.~Kim, ``Dyser: Unifying functionality and parallelism specialization for energy-efficient computing,'' {\em IEEE Micro}, vol.~32, no.~5, 2012.

\bibitem{srinath2014architectural}
S.~Srinath, B.~Ilbeyi, M.~Tan, G.~Liu, Z.~Zhang, and C.~Batten, ``Architectural specialization for inter-iteration loop dependence patterns,'' in {\em 2014 47th Annual IEEE/ACM Intl. Symp. on Microarchitecture}, IEEE, 2014.

\bibitem{gupta2011bundled}
S.~Gupta, S.~Feng, A.~Ansari, S.~Mahlke, and D.~August, ``Bundled execution of recurring traces for energy-efficient general purpose processing,'' in {\em Proceedings of the 44th Annual IEEE/ACM Intl. Symp. on Microarchitecture}, 2011.

\bibitem{vachharajani2007speculative}
N.~Vachharajani, R.~Rangan, E.~Raman, M.~J. Bridges, G.~Ottoni, and D.~I. August, ``Speculative decoupled software pipelining,'' in {\em 16th Intl. Conf. on Parallel Architecture and Compilation Techniques (PACT 2007)}, IEEE, 2007.

\bibitem{voitsechov2015control}
D.~Voitsechov and Y.~Etsion, ``Control flow coalescing on a hybrid dataflow/von neumann gpgpu,'' in {\em Proceedings of the 48th Intl. Symp. on Microarchitecture}, 2015.

\bibitem{voitsechov2014single}
D.~Voitsechov and Y.~Etsion, ``Single-graph multiple flows: Energy efficient design alternative for gpgpus,'' {\em ACM SIGARCH computer architecture news}, vol.~42, no.~3, 2014.

\bibitem{voitsechov2018inter}
D.~Voitsechov, O.~Port, and Y.~Etsion, ``Inter-thread communication in multithreaded, reconfigurable coarse-grain arrays,'' in {\em 2018 51st Annual IEEE/ACM Intl. Symp. on Microarchitecture (MICRO)}, IEEE, 2018.

\bibitem{SPEC17}
``Spec cpu 2017.'' \url{https://www.spec.org/cpu2017/}.

\bibitem{Intel_pin}
``Intel pin - a dynamic binary instrumentation tool.'' \url{https://www.intel.com/content/www/us/en/developer/articles/tool/pin-a-binary-instrumentation-tool-downloads.html}.

\bibitem{luk2005pin}
C.-K. Luk, R.~Cohn, R.~Muth, H.~Patil, A.~Klauser, G.~Lowney, S.~Wallace, V.~J. Reddi, and K.~Hazelwood, ``Pin: building customized program analysis tools with dynamic instrumentation,'' {\em Acm sigplan notices}, vol.~40, no.~6, 2005.

\bibitem{hauser1997garp}
J.~R. Hauser and J.~Wawrzynek, ``Garp: A mips processor with a reconfigurable coprocessor,'' in {\em Proceedings. The 5th Annual IEEE Symp. on Field-Programmable Custom Computing Machines Cat. No. 97TB100186)}, IEEE, 1997.

\bibitem{callahan2000garp}
T.~J. Callahan, J.~R. Hauser, and J.~Wawrzynek, ``The garp architecture and c compiler,'' {\em Computer}, vol.~33, no.~4, 2000.

\bibitem{callahan2000adapting}
T.~J. Callahan and J.~Wawrzynek, ``Adapting software pipelining for reconfigurable computing,'' in {\em Proceedings of the 2000 Intl. Conf. on Compilers, architecture, and synthesis for embedded systems}, 2000.

\bibitem{fan2009bridging}
K.~Fan, M.~Kudlur, G.~Dasika, and S.~Mahlke, ``Bridging the computation gap between programmable processors and hardwired accelerators,'' in {\em 2009 IEEE 15th Intl. Symp. on High Performance Computer Architecture}, IEEE, 2009.

\bibitem{fan2005cost}
K.~Fan, M.~Kudlur, H.~Park, and S.~Mahlke, ``Cost sensitive modulo scheduling in a loop accelerator synthesis system,'' in {\em 38th Annual IEEE/ACM Intl. Symp. on Microarchitecture (MICRO'05)}, IEEE, 2005.

\end{thebibliography}

\end{document}